\documentclass[letterpaper,twoside,sort&compress,super,3p,12pt]{elsarticle-BMJ}
\usepackage{fancyhdr}
\usepackage{graphicx}
\usepackage{amsmath,amssymb,mathtools}
\usepackage{theorem}
\usepackage{epstopdf}
\ifpdf
  \DeclareGraphicsExtensions{.eps,.pdf,.png,.jpg}
\else
  \DeclareGraphicsExtensions{.eps}
\fi
\usepackage[table]{xcolor} 
\usepackage{color}
\usepackage[flushleft]{threeparttable} 
\usepackage{multirow, fancybox}
\usepackage{booktabs}
\usepackage{pdflscape} 
\usepackage{paralist} 

\usepackage{url}
\usepackage{cleveref}

\newtheorem{theorem}{Theorem}
\theoremstyle{definition}

\biboptions{comma}
\makeatletter
\renewcommand\@biblabel[1]{#1} 
\makeatother
\usepackage{multicol,array}
\usepackage{tabularx}
\usepackage[normalem]{ulem}
\usepackage{arydshln}
\usepackage{siunitx} 
\allowdisplaybreaks
\def\W{\textit{Wolbachia}}
\def\Ws{\textit{Wolbachia~}}
\def\An{\textit{Anopheles~}}
\def\Ae{\textit{Aedes~}}

\newif\ifnotesw \noteswtrue

\newcommand{\bbG}{{\mathbb{G}}}
\newcommand{\bbM}{{\mathbb{M}}}
\newcommand{\bbR}{{\mathbb{R}}}
\newcommand{\bbk}{{\mathbb{K}}}

\newcommand{\bF}{{\boldsymbol{F}}}
\newcommand{\bV}{{\boldsymbol{V}}}
\newcommand{\bJ}{{\boldsymbol{J}}}
\newcommand{\bX}{{\boldsymbol{X}}}

\newcommand\scalemath[2]{\scalebox{#1}{\mbox{\ensuremath{\displaystyle #2}}}}

\begin{document}
\begin{frontmatter}
\title{Modeling sustained transmission of  \textit{Wolbachia} among \textit{Anopheles} mosquitoes: Implications for malaria control in Haiti}

\author[Tmath]{Daniela Florez}
\author[Tph]{Alyssa Young}
\author[Tph]{Kerlly J. Bernab\'e}
\author[Tmath]{James M. Hyman}
\author[UTSA]{Zhuolin Qu*}
\ead{zhuolin.qu@utsa.edu}
\address[Tmath]{\raggedright Department of Mathematics, Tulane University, New Orleans, LA, USA}
\address[Tph]{School of Public Health and Tropical Medicine, Tulane University, New Orleans, LA, USA}
\address[UTSA]{Department of Mathematics, University of Texas at San Antonio, San Antonio, TX, USA}

\begin{abstract}
\Ws infection in \textit{Anopheles albimanus} mosquitoes can render mosquitoes less capable of spreading malaria. We develop and analyze an ordinary differential equation model to evaluate the effectiveness of \W-based vector control strategies among wild \An mosquitoes in Haiti. The model tracks the mosquito life stages, including egg, larva, and adult (male and female). It also accounts for critical biological effects, such as the maternal transmission of \Ws through infected females and cytoplasmic incompatibility, which effectively sterilizes uninfected females when they mate with infected males. We derived and interpreted dimensionless numbers, including the basic reproductive number and next-generation numbers. The proposed system presents backward bifurcation, which indicates a threshold infection that needs to be exceeded to establish a stable \Ws infection. The sensitivity analysis ranks the relative importance of the epidemiological parameters at the baseline. We simulate different intervention scenarios, including pre-release mitigation using larviciding and thermal fogging before the release, multiple releases of infected populations, and different release timing. Our simulations show that the most efficient approach to establishing \Ws is to release all the infected mosquitoes immediately after the pre-release mitigation process. Also, the model predicts that it is more efficient to release during the dry season than the wet season.} 


\end{abstract}

\end{frontmatter} 
\thispagestyle{fancy}
\newpage

\section{Introduction}
Malaria is a febrile illness caused by several species of \textit{Plasmodium} protozoan parasites through the bite of an infected female \textit{Anopheles (An.)} mosquito.  
Falciparum malaria is a leading cause of death globally and the most lethal of the five known species of \textit{Plasmodium} that can infect humans.
Current efforts to control malaria typically focus on strengthening surveillance, administering seasonal malaria chemoprophylaxis, or reducing mosquito populations, through means such as distributing insecticide-treated bed nets (ITNs), implementing larval control, and/or conducting indoor residual spraying (IRS). Due to increasing insecticide resistance, the impact of climate change, and other environmental factors on mosquito breeding and feeding behavior, more sustainable and effective mitigation strategies are necessary. 

\textit{Wolbachia pipientis} is a gram-negative, intracellular endosymbiotic bacterium that naturally infects over 75\% of all arthropods \cite{pan2017wolbachia, mustafa2016wolbachia}, including mosquitoes that spread human diseases. Transinfection of \textit{Aedes sp.} mosquitoes shown to be effective at controlling Dengue fever, Chikungunya, and Zika virus transmission. Recently, evidence suggests that similar approaches can be used to control the spread of \textit{P. falciparum} malaria.

\Ws has been used as a \emph{population suppression} strategy by releasing infected males. Evidence suggests that the presence of the bacteria modifies the paternal chromosomes during spermatogenesis \cite{hancock2011strategies}. When mating with an uninfected female, the sperm of the infected male cannot form viable offspring during the egg fertilization process, resulting in unhatched eggs \cite{gomes2018infection}. This \W-induced cytoplasmic incompatibility (CI) phenomenon provides an alternative approach similar to adulticides. This population suppression strategy requires the continual introduction of infected males to sustain the suppression, and it has the sustainability issue as traditional approaches. When the intervention stops, mosquito populations may re-emerge. Moreover, the strategy can be hard to deploy in practice due to the accidental release of infected females, which may produce infected offspring and undermine the process \cite{zheng2019incompatible,zhang2015combining}.  

\Ws has also been used as a \emph{population replacement} strategy (\cref{tab:CI}) for controlling diseases by releasing both male and female infected mosquitoes into the field to replace the wild population \cite{walker2011wmel,hoffmann2011successful}. In some \An species, \Ws reduces the number of \textit{P. falciparum} oocysts and sporozoites \cite{bian2013wolbachia}, and \W-infected \An mosquitoes are less effective in transmitting the parasite. Passing of such anti-pathogenic traits to offspring is achievable as \Ws exhibits high rates of maternal transmission in both \Ae and \An \textit{sp.} mosquitoes \cite{pan2017wolbachia,joshi2014wolbachia, joshi2017maternally}. This leads to a population replacement strategy: instead of removing the wild mosquitoes, the goal is to infect mosquitoes with \Ws and replace the wild mosquito population with the infected ones that can no longer transmit the malaria parasite. Field  studies show that the population replacement strategy can be a more sustainable approach \cite{hoffmann2014stability,oneill2018use}. 

\begin{table}[htbp] 
\centering
\caption{Role of \Ws in mosquito population replacement versus population suppression \label{tab:CI}}
\begin{tabular}{@{}p{1.5cm}p{6cm}p{6cm}@{}}
\toprule
& Population Replacement & Population Suppression\\
\midrule
\textbf{Goal} & Replace wild mosquito population with \W-infected mosquitoes that have significantly lower competence and cannot transmit parasite as efficiently & Introduce male mosquitoes that cannot produce viable offspring, which limits the ability of the mosquito to reproduce and reduces mosquito population\\
\midrule
\textbf{Role of CI} & Infected females can mate successfully with infected males providing them with an evolutionary advantage over uninfected females & The sperm of the infected male is unable to form viable offspring during the egg fertilization process, and as a result, eggs do not hatch\\
\midrule
\textbf{Release} & Release infected males and females & Release infected males only\\
\bottomrule
\end{tabular}
\end{table}

Infecting \An mosquitoes with \textit{wMelPop} and \textit{wAlbB} strains of \Ws show a reduction in \textit{P. falciparum} sporozoite and oocyst levels in specific species (\cref{tab:ACI}), and the \textit{wAlbB} strain exhibits perfect maternal (vertical) transmission in \textit{An. stephensi} \cite{joshi2014wolbachia, joshi2017maternally}. Since almost all offspring of \textit{wAlbB}-infected females will be infected, this specific strain is a reasonable choice to create a sustained population of \W-infected wild \An  mosquitoes. Our focus is to identify population replacement strategies for creating a sustained \textit{wAlbB} infection within a population of wild \An mosquitoes. 

\begin{table}[t] 
\begin{center}
\caption{\Ws strains, \textit{\An} species, and corresponding impact on vector and \textit{P. falciparum} parasite replication.\label{tab:ACI}}
\begin{tabular}{@{}p{2cm}p{2cm}p{4cm}p{4cm}p{1.5cm}@{}}
\toprule
\textit{Wolbachia} strain & \textit{Anopheles} species  & Impact on vector &Impact on \newline \textit{P. falciparum} & Reference \\
\midrule
\multirow{5}{*}{$wAnga$} & \textit{coluzzii}   & No CI, \newline increases egg laying rate & Reduces sporozoite prevalence  & \cite{shaw2016wolbachia, childs2020role, sicard2019wolbachia, gomes2017} \\
& \textit{funestus}    & No CI  & Unknown &  \cite{sicard2019wolbachia} \\ 
& \textit{gambiae}     & No CI & Unknown  & \cite{sicard2019wolbachia} \\ 
& \textit{arabiensis}     & No CI & Unknown & \cite{sicard2019wolbachia} \\
\midrule
$wAlbB$ & \textit{stephensi}   & Almost complete CI, reduces egg hatching rate, perfect maternal transmission, no impact on female lifespan & Reduces sporozoite and oocyst levels &  \cite{joshi2014wolbachia, bian2013wolbachia} \\
\midrule
$wPip$& \textit{gambiae}     & CI, reduces egg development rate & Unknown &  \cite{adams2021wolbachia} \\
\midrule
$wMelPop$ & \textit{gambiae}     & No effect on lifespan & Significantly reduces oocyst level &  \cite{hughes2011wolbachia} \\
\bottomrule
\end{tabular}
\end{center}
\end{table}

Mathematical models for studying the \Ws infection in mosquitoes primarily focus on arboviruses spread by \textit{Aedes spp.} mosquitoes. Xue et al. \cite{xue2018comparing} compared the impact of infecting \textit{Aedes aegypti} and \textit{Aedes albopictus} mosquitoes with \textit{wAlbB} and \textit{wMel} in reducing the transmission of Dengue, Chikungunya, and Zika viruses. 
This study analyzed a system of seven ordinary differential equations (ODEs) that accounted for the reduced fitness of \textit{Wolbachia-}infected mosquitoes, reduced transmissibility of infected mosquitoes, and behavior changes of infected humans caused by disease. This model was based on previous studies that modeled the potential of establishing a population of wild \W-infected \Ae mosquitoes \cite{qu2018modeling, qu2019generating, Xue2017Two} and incorporated a series of two-sex compartmental models for \Ws transmission in \Ae mosquitoes. These models quantify the effectiveness of different approaches to ensure the sustained transmission of \Ws within wild \textit{Aedes} mosquitoes.

These studies on non-malaria-transmitting mosquitoes formed the basis of our \An mosquito model. We extended the models by subdividing the aquatic stage into the egg and larval/pupae stages. We assume the time it takes for a newly born mosquito to be impregnated is small compared to its lifetime; therefore, we eliminated the impregnated mosquito compartments \cite{qu2018modeling}.  

We chose malaria-specific interventions conducted before the release of infected \Ws infection. These interventions are larvicides, ultra-low volume spraying, and thermal fogging, although these interventions are not necessarily implemented in Haiti. These interventions can accelerate establishing an endemic infection of \Ws among \An mosquitoes. We simulate the combination of these vector control strategies for one release scenario of \W-infected mosquitoes. 

We found that enough \W-infected mosquitoes need to be released so the fraction of infected mosquitoes exceeds $34\%$, which corresponds to a threshold condition for \Ws to spread in wild \An populations. Otherwise, \Ws infection dies out in the wild population. 
Using larvicides and thermal fogging before releasing \W-infected mosquitoes reduces the wild mosquito population and, thus, reduces the number of infected mosquitoes that need to be released. Or, if the same number of infected mosquitoes are released, pre-release mitigation accelerates the establishment of endemic \Ws infection among wild \An. 

We parameterized our model based on remote sensing data extracted from Grand'Anse, Haiti. This seasonal data is crucial to evaluate the impact of variations in the wild mosquito population on establishing a stable, high infection of \Ws among wild mosquitoes. 
We compared the effectiveness of releasing \W-infected mosquitoes during different seasons. 
The population of wild mosquitoes is the smallest during the dry season, and far fewer infected mosquitoes are needed to exceed the threshold. 

After defining our compartmental ODE model (\cref{sec:model}), we analyze the model by introducing the next-generation numbers, $\bbG_{0u}$ and $\bbG_{0w}$, for the uninfected and infected mosquito populations (\cref{sec:next}). We derive the reproductive number, $\bbR_0$, for the spread of \Ws in the mosquito population and illustrate how $\bbR_0$ can be interpreted in terms of the next-generation numbers (\cref{sec:R0}). We then compare different release scenarios and investigate the effect of concurrent malaria vector control interventions and the impact of seasonality (\cref{sec:numerical}). 


\section{Model Description}\label{sec:model}
\begin{figure}[htbp]
\centering
\includegraphics[width=.7\textwidth]{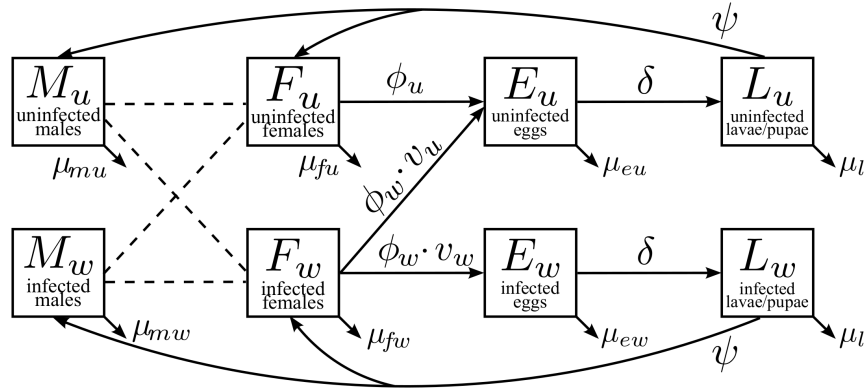}
\caption{Maternal transmission of \Ws in mosquitoes. The adult population of males and females is divided into compartments based on the infection status. Uninfected females ($F_{u}$) produce uninfected eggs ($E_u$) at an egg-laying rate of $\phi_u$. Infected females ($F_w$) produce a fraction of $v_w$ infected eggs ($E_w$) with a rate of $\phi_w$. Then, eggs develop into the larval stage at a hatching rate of $\delta$. Larval-stage mosquitoes emerge at a rate of $\psi$ and develop into adult mosquitoes. Death rates at different stages are denoted by $\mu_*$.}
\label{fig:block diagram}
\end{figure}

Our multi-stage, two-sex model partitions the mosquito population by life stages and \W-infection status. \Cref{fig:block diagram} illustrates the maternal transmission of \textit{wAlbB} in mosquitoes. The adult stages include uninfected males ($M_u$), infected males ($M_w$), uninfected females ($F_u$), and infected females ($F_w$). The uninfected eggs ($E_u$) and infected eggs ($E_w$) are separate compartments. The larvae and pupae stages were combined into one compartment for the uninfected ($L_u$) and infected groups ($L_w$). Model parameters and baseline values for the simulations are shown in \cref{tab:param}. The details on the parameter estimates are found in section 4.1.

\paragraph{\textbf{Male adult mosquitoes ($M_u$ and $M_w$)}}
Uninfected and infected males have mean lifespans of $\tau_{mu}$ or $\tau_{mw}$. We assume an exponential survival rate, which leads to constant daily death rates of $\mu_{mu}=1/\tau_{mu}$ or $\mu_{mw}=1/\tau_{mw}$. 
The uninfected and infected males randomly mix and impregnate females. We assume that \W-infection minimally affects the mating behavior of mosquitoes. Therefore, infected males are nearly as competent as the uninfected males \cite{joshi2014wolbachia}.
The probability that a random male mosquito is uninfected or infected is determined by the proportions 
\begin{equation}\label{eq:MM}
\bbM_u = \frac{M_u}{M_u + M_w}, \text{~and~}
\bbM_w = 1- \bbM_u = \frac{M_w}{M_u + M_w}~.
\end{equation}

\paragraph{\textbf{Female adult mosquitoes ($F_u$ and $F_w$)}} \W-infection may affect the lifespan of females \cite{joshi2014wolbachia}. The uninfected and infected females have a mean lifespan of $\tau_{fu}$ and $\tau_{fw}$, respectively and gives constant daily death rates of $\mu_{fu} = 1/\tau_{fu}$ and $\mu_{fw} = 1/\tau_{fw}$. 

For simplification, we assume that all females are impregnated soon after hatching and do not distinguish between nonpregnant and pregnant females (see \cite{qu2019generating,Xue2017Two}). This assumption slightly adjusts the average daily egg-laying rates $\phi_u$ and $\phi_w$ for uninfected and infected females, respectively.

\paragraph{\textbf{Egg stages ($E_u$ and $E_w$)}}  
The fraction of eggs produced by infected females infected ($v_w$) is independent of the infection status of males.   This fraction is known as the maternal transmission rate. The remainder of the eggs are uninfected ($v_u = 1-v_w$).

When an uninfected female is impregnated by an infected male ($F_u$ cross $M_w$, with probability $\bbM_w$, as defined in \cref{eq:MM}), the \W-induced CI may cause a fraction of the impregnated females to lay nonviable eggs ($c_i$). Thus, a fraction $c_i \bbM_w$ of uninfected females are sterile, and the remainder of the eggs are fertile and produce viable uninfected eggs ($(1-c_i)\bbM_w$). Thus, the birth rate of the uninfected eggs is given by 
$$
\bbM_u \phi_u +\bbM_w (1-c_i) \phi_u  +\bbM_w c_i \cdot 0~. 
$$
Without \W-induced CI ($c_i=0$), the birth rate of the viable uninfected is $\phi_u$ per day. For the baseline simulations with $wAlbB$, we assume a complete CI ($c_i=1$), and the birth rate for $E_u$ is $ \phi_u \bbM_u$.

Although \W-infection may not affect the total number of offspring that the infected female reproduces \cite{joshi2014wolbachia}, it can impact the survivorship of the eggs produced. Thus, we assume that the uninfected and infected eggs have daily death rates of $\mu_{eu}$ and $\mu_{ew}$. Also, we assume surviving eggs then hatch at a rate of $\delta$ regardless of infection status.


\paragraph{\textbf{Larvae/pupae stages ($L_u$ and $L_w$)}} 
We combined the larvae and pupae stages and limited the population using a logistic carrying capacity constraint ($K_l$). The carrying capacity is dependent on the availability of water and food resources, and it is incorporated into the model by applying the constraint
\begin{equation}\label{eq:CC}
\bbk = 1-\frac{L_u + L_w}{K_l}
\end{equation}
on the birth rate of the larvae/pupae group. The carrying capacity accounts for the seasonal variations that affect mosquito populations. We use the seasonally adjusted carrying capacity to investigate the impact of seasonality on establishing \Ws by releasing infected mosquitoes (\cref{sec:seasonality}). 

The adult mosquitoes emerge from the larvae stages at a constant rate ($\psi$). The emergence rates are not significantly different between the wild and infected cohort \cite{joshi2014wolbachia}. The fraction of larvae that emerge to become females is denoted as $b_f$. The fraction of larvae that emerge to become males is defined as $b_m=1-b_f$.

These assumptions are satisfied by the solution to the system of differential equations
\begin{equation}
\begin{aligned}\label{eq:VH-ODE}
\frac{dM_u}{dt}&= b_m\psi L_u - \mu_{mu}M_u~,\\
\frac{dM_w}{dt}&= b_m \psi L_w - \mu_{mw} M_w~,\\
\frac{dF_{u}}{dt}&= b_f \psi L_u -  \mu_{fu} F_u~,\\
\frac{dF_{w}}{dt}&=b_f\psi L_w - \mu_{fw}F_w~,\\
\frac{dE_u}{dt}&= \phi_u\bbM_u F_u + \phi_wv_uF_w - \delta E_u - \mu_{eu} E_u~, \\
\frac{dE_w}{dt}&= v_w \phi_wF_w - \delta E_w - \mu_{ew} E_w~,\\
\frac{dL_u}{dt}&= \delta \bbk  E_u - \psi L_u - \mu_l L_u~,\\
\frac{dL_w}{dt}&= \delta  \bbk  E_w - \psi L_w - \mu_l L_w~.
\end{aligned}
\end{equation}
Here $\bbM_u$ and $\bbk$ are dimensionless quantities, defined as nonlinear functions of the state variables (\cref{eq:MM,eq:CC}). 

\begin{table}[htbp]
\centering
\caption{Model parameters and the baseline values. All rates have the unit day$^{-1}$. \label{tab:param}}
\begin{tabular}{@{}clrc@{}}
\toprule
& Description   & Value &  Reference \\
\midrule
&{\bf specific to \textit{Anopheles spp.}}&&\\
$\delta $ & Hatching rate  for  eggs ($=1/\tau_{\delta}$) & 1/3  & \cite{CDCMalaria}\\
$\psi$ & Emergence rate  for larvae  ($=1/\tau_{\psi}$)  &  1/18 &  \cite{joshi2014wolbachia} \\
$\mu_{fu}$ & Death rate for uninfected females ($=1/\tau_{fu}$) &  1/13 &   \cite{joshi2014wolbachia,CDCMalaria}\\
$\mu_{fw}$ & Death rate for infected females ($=1/\tau_{fw}$)  &  1/15 & \cite{joshi2014wolbachia,CDCMalaria} \\
$\mu_{mu}$ & Death rate for uninfected males ($=1/\tau_{mu}$) & 1/7  & \cite{joshi2014wolbachia,CDCMalaria}  \\
$\mu_{mw}$ & Death rate for infected males ($=1/\tau_{mw}$) & 1/7 & \cite{joshi2014wolbachia,CDCMalaria}  \\
$ \mu_{eu}$ & Death rate for uninfected eggs & 0.12 & \cite{joshi2014wolbachia}\\
$\mu_{ew}$ & Death rate for infected eggs   & 0.33 & \cite{joshi2014wolbachia}\\
$\mu_l$ & Death rate for larvae & 0.01 & \cite{joshi2014wolbachia}\\
$\phi_u$  & Per capita egg laying rate for wild females & 3.8 & \cite{joshi2014wolbachia} \\
$\phi_w$ & Per capita egg laying rate for infected females & 3.3 & \cite{joshi2014wolbachia}\\
$v_w$ & \textit{wAlbB} maternal transmission fraction & 1 & \cite{bian2013wolbachia} \\
$c_i$ & \textit{wAlbB} CI fraction &  1 &  \\
&{\bf not specific to \textit{Anopheles spp.}}&&\\
$b_f$ & Fraction of larvae emerge to females  & 0.5  & \cite{tun2000effects} \\
$b_m$ & Fraction of larvae emerge to males  & 0.5  & \cite{tun2000effects} \\
$K_l$ & Carrying capacity of larvae/pupae stages & $2\times{10}^5$ & Assume \\
\bottomrule
\end{tabular}
\end{table}

\section{Model Analysis}
We analyze the model \cref{eq:VH-ODE} by first defining two \textit{next-generation numbers}, $\bbG_{0u}$ and $\bbG_{0w}$. These factors provide insightful information on mosquito reproduction and reflect the competition between the uninfected and infected cohorts during the population replacement process \cite{qu2018modeling}.

\subsection{Next-Generation Numbers}\label{sec:next}
When there is no \Ws infection in the population, the average number of uninfected eggs that an uninfected female lays over a lifetime is given by $\phi_u/\mu_{fu}$. A fraction $\delta/(\delta+\mu_{eu})$ of these uninfected eggs can survive and develop into the larvae/pupae stage. With probability $\psi/(\psi+\mu_l)$, larvae develop into uninfected female adults. The product of these factors gives the number of new uninfected females generated by one uninfected female through one generation,
\begin{equation}
\mathbb{G}_{0u}=b_f \frac{\psi}{\psi+\mu_l}\, \frac{\delta}{\delta+\mu_{eu}} \,\frac{\phi_u}{\mu_{fu}}~,
\label{G0u}
\end{equation}
which we define as the next-generation number for the uninfected population. Near the baseline parameter values (\cref{tab:param}), we have $\bbG_{0u}>1$, indicating that the wild mosquito population can persist when there are no \W-infected mosquitoes. 

Similarly, we define the next-generation number for the infected population,
\begin{equation}\label{G0w}
\mathbb{G}_{0w}=v_wb_f \frac{\psi}{\psi+\mu_l}\, \frac{\delta}{\delta+\mu_{ew}} \,\frac{\phi_w}{\mu_{fw}}~,
\end{equation}
where $v_w$ is the maternal transmission rate, which gives the fraction of infected eggs produced by the \W-infected females.

\subsection{Equilibria and Basic Reproductive Number}\label{sec:R0}
The model has three types of equilibrium points: disease-free equilibrium, complete-infection equilibrium, and endemic equilibrium. 
\subsubsection{Disease-free equilibrium (DFE)}
We derive the DFE by setting the populations in all infected stages equal to zero in the system \cref{eq:VH-ODE} ($E_w=L_w=F_w=M_w=0$). The corresponding equilibrium solution gives the DFE, denoted by $\bX^0=(E_u^0,0,L_u^0,0, F_u^0,0,M_u^0,0)^T$.
\begin{align}
E_u^0&=b_f \frac{\psi}{\mu_{fu}} \frac{\phi_u}{\delta+\mu_{eu}} L_u^0~,\nonumber\\
L_u^0&=K_l\bigg(1-\frac{1}{\bbG_{0u}}\bigg)~,\label{eq:DFE}\\
F_u^0&=b_f \frac{\psi}{\mu_{fu}} L_u^0~,\nonumber\\
M_u^0&=b_m \frac{ \psi}{\mu_{mu}} L_u^0~\nonumber,
\end{align}
and $\bbG_{0u}$ is the next-generation number for the uninfected population (\cref{G0u}).

\subsubsection{Complete-infection Equilibrium (CIE)}
The CIE exists when assuming perfect maternal transmission, that is, $v_w=1$, and all the mosquitoes are infected. We derive CIE by setting all the uninfected compartments to zero, i.e., $E_u=L_u=F_u=M_u=0$, in the system \cref{eq:VH-ODE}, and the corresponding equilibrium solution gives the CIE, which is denoted by $\bX^c=(0,E_w^c,0,L_w^c,0,F_w^c, 0, M_w^c)^T$, 
\begin{align}
E_w^c&=b_f\frac{\phi_w}{\delta+\mu_{ew}}\,\frac{\psi}{\mu_{fw}}L_w^c~,\nonumber\\
L_w^c&=K_l\bigg(1-\frac{1}{\bbG_{0w}}\bigg)~,\label{eq:CIE}\\
F_w^c&=b_f\frac{\psi}{\mu_{fw}}L_w^c~,\nonumber\\
M_w^c&=b_m\frac{\psi}{\mu_{mw}}L_w^c~\nonumber,
\end{align}
and $\bbG_{0w}$ is the next-generation number for the infected population (\cref{G0w}).

\subsubsection{Basic Reproductive Number} 
After deriving the basic reproductive number by following the classical theory of the next-generation matrix, we interpret the obtained expression from a biological perspective.
\paragraph{\textbf{Derivation using the next-generation Matrix}}
Consider the infected compartments in the model \cref{eq:VH-ODE}, denoted by $\bX_w$ = ($E_w$, $L_w$, $F_w$, $M_w$)$^T$, and define a subsystem for these variables,
\begin{equation*}
\frac{d\bX_w}{dt} = \frac{d}{dt}\begin{pmatrix}
E_w \\
L_w \\
F_w \\ 
M_w
\end{pmatrix}
= \bF- \bV =\begin{pmatrix}
v_w\phi_wF_w  \\
0 \\ 
0 \\ 
0
\end{pmatrix} 
-\begin{pmatrix}
(\delta+\mu_{ew})E_w \\
-\delta(1-\frac{L_u + L_w}{K_l})E_w+(\psi+\mu_l)L_w \\
-b_f\psi L_w + \mu_{f_w}F_w \\
-b_m\psi L_w+ \mu_{mw}M_w
\end{pmatrix}~,
\end{equation*}
where the vectors $\bF$ and $\bV$ represent the rate of new infections and the transition rate among the infected compartments. We then linearized the equation at the DFE and obtained the Jacobian matrices $\bJ_\bF(\bX^0)$ and $\bJ_\bV(\bX^0)$,
\begin{equation*}
\bJ_\bF = 
\begin{pmatrix}
0 & 0 & v_w\phi_w & 0 \\
0 & 0 & 0 & 0 \\
0 & 0 & 0 & 0 \\
0 & 0 & 0 & 0 \\
\end{pmatrix}~,\quad
\bJ_\bV= 
\begin{pmatrix}
\delta+\mu_{ew} & 0 & 0 & 0 \\
-\delta/\bbG_{0u} &\psi +\mu_l & 0 & 0 \\
0& -b_f\psi  & \mu_{fw} & 0 \\
0 & -b_m \psi & 0 & \mu_{mw}
\end{pmatrix}~,
\end{equation*}
and the basic reproductive number is given by 
\begin{equation}
\bbR_0:= \text{spectral radius of~} \bJ_\bF \bJ_\bV^{-1}=v_w \frac{\mu_{fu}\phi_w(\delta+\mu_{eu})}{\mu_{fw}\phi_u (\delta +\mu_{ew})}~.
\end{equation}

\paragraph{\textbf{Interpretation of the basic reproductive number}}
The basic reproductive number derived can be written as the ratio of the next-generation numbers
\begin{equation*}
\bbR_0 = \frac{\bbG_{0w}}{\bbG_{0u}} = 
v_w \frac{\mu_{fu}\phi_w(\delta+\mu_{eu})}{\mu_{fw}\phi_u (\delta +\mu_{ew})}~.
\end{equation*}
Recall the definition of the next-generation numbers $\bbG_{0u}$ and $\bbG_{0w}$ (\cref{sec:next}), which represent the numbers of new offspring reproduced per generation among the uninfected and infected cohorts, assuming the system is near DFE. The ratio of these next-generation numbers is an estimate of the average number of infected offspring generated per infected individual at the DFE.

When $\bbG_{0w}>\bbG_{0u}$ or $\bbR_0>1$, the infected mosquitoes reproduce than the uninfected ones. Therefore, the small infection will spread in the population. The infected population experiences a fitness cost, and $\bbR_0<1$ is the practical case at baseline. Hence, the naturally uninfected population will wipe out a small introduction of an infected population. Consequently, the ratio of next-generation numbers is a threshold condition for a small initial invasion of the \W-infected population in the wild mosquito population.

\subsubsection{Endemic Equilibrium (EE)}
During imperfect maternal transmission, $v_w <1$, some infected females produce uninfected offspring, and the CIE cannot be maintained. Instead, there is EE, where infected and uninfected mosquitoes coexist.

We first define $r_{wu}$ as the ratio between the infected and uninfected larvae/pupae stages (i.e., $r_{wu}=L_w/L_u$) which is a key dimensionless quantity in the derivation. Then EE, denoted by $\bX^*=(M_u^*,M_w^*,F_u^*,F_w^*,E_u^*,E_w^*,L_u^*,L_w^*)^T$, can be written in terms of $r_{wu}$ as follows:
\begin{align}
M_u^*&=b_m\frac{\psi}{\mu_{mu}}L_u^*~,\nonumber\\
M_w^*&=r_{wu}  b_m\frac{\psi}{\mu_{mw}}L_u^*~,\nonumber\\
F_u^*&= b_f\frac{\psi}{\mu_{fu}}L_u^*,\nonumber\\
F_w^*&= r_{wu} b_f\frac{\psi}{\mu_{fw}}L_u^*~,\label{eq:EE}\\
E_u^*&=\frac{b_f\psi}{\delta+\mu_{eu}}\left(\frac{\phi_{u}}{\mu_{fu}}\frac{1}{1+r_{wu}}+\frac{v_u\phi_w}{\mu_{fw}}r_{wu}\right)L_u^*~,\nonumber\\
E_w^*&=r_{wu} \frac{b_f\psi}{\delta+\mu_{ew}}\frac{v_w\phi_w}{\mu_{fw}}L_u^*=r_{wu}E_u^*~,\nonumber\\
L_u^*&=\frac{1}{1+r_{wu}}K_l\bigg(1-\frac{1}{\bbG_{0w}}\bigg)~,\nonumber\\
L_w^*&=r_{wu}L_u^*~\nonumber
\end{align}
We assume the death rate for uninfected males and infected males are the same ($\mu_{mu}=\mu_{mw}$). The ratio $r_{wu}$ satisfies the  quadratic relation that involves the maternal transmission rate, infection leakage rate, and basic reproductive number,
\begin{equation}\label{eq:r_wu relation}
\frac{v_u}{v_w}\frac{\delta+\mu_{ew}}{\delta+\mu_{eu}}r_{wu}^2+\left(\frac{v_u}{v_w}\frac{\delta+\mu_{ew}}{\delta+\mu_{eu}}-1\right)r_{wu}+\frac{1-\bbR_0}{\bbR_0}=0~.
\end{equation}
Under the special case of perfect maternal transmission ($v_w=1$), \cref{eq:r_wu relation} degenerates to a linear relation,
\begin{equation*}
r_{wu}^*=\frac{L_w^*}{L_u^*}=\frac{1-\bbR_0}{\bbR_0}~.
\end{equation*}
To have a physically relevant endemic equilibrium, we need to impose $r_{wu}^*>0$. This implies that $\bbR_0$ must be between 0 and 1. Our baseline estimate for $\bbR_0$ is 0.68.

\subsection{Stability and Bifurcation Analysis}
The stability analysis of the equilibrium points helps to characterize the solution dynamics, and it indicates a threshold condition for establishing stable \W-infection among mosquitoes. 

The stability of an equilibrium is determined by the signs of the eigenvalues of the Jacobian matrix for the system of ODEs linearized about the equilibrium. We present the conclusions on the stability analysis below. The proofs of \Cref{thm:stabDFE} and \Cref{thm:stabCIE}  are in \cref{sec:A1,sec:A2}. We numerically verify the conjecture \Cref{thm:stabEE}. These conclusions are comparable to the ones in \cite{qu2018modeling} for a similar model structure.
\vspace{-0.3cm}
\begin{theorem}[Stability of the Disease-free Equilibrium]\label{thm:stabDFE}
The DFE (\cref{eq:DFE}) of system \cref{eq:VH-ODE} is locally asymptotically stable provided that $\mathbb{G}_{0u}>1$ and $\mathbb{R}_{0}<1$. 
\end{theorem}
\vspace{-0.3cm}
\begin{theorem}[Stability of the Complete Infection Equilibrium]\label{thm:stabCIE}
The CIE (\cref{eq:CIE}) of the system  \cref{eq:VH-ODE}  is locally asymptotically stable provided that $\mathbb{G}_{0w}>1$.
\end{theorem}
\vspace{-0.3cm}
\begin{theorem}[Stability of the Endemic Equilibrium]\label{thm:stabEE} When having the perfect maternal transmission ($v_w=1$), the physically relevant EE (\cref{eq:EE}) of the system  \cref{eq:VH-ODE} exists for $\bbR_0<1$ and $\bbG_{0w}>1$, and it is an unstable equilibria.
\end{theorem}

\begin{figure}[htbp]
\centering
\includegraphics[width=0.6\linewidth]{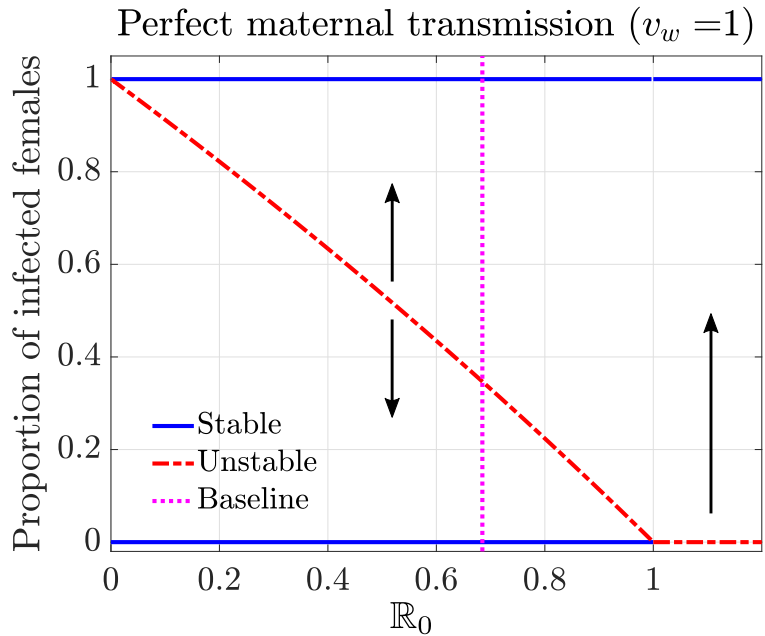} 
\caption{Bifurcation diagram characterizing the threshold condition for establishing a stable infection in mosquitoes, given a perfect maternal transmission rate ($v_w=1$). The solid blue curves represent the stable equilibrium. The red dashed curve corresponds to the unstable equilibrium, which serves as the threshold condition. At the baseline case (vertical dotted line, $\bbR_0=0.68$), the threshold infection rate among females is $0.34$.}
\label{fig:bifurcation}
\end{figure}

Based on the conclusions above, we generated the bifurcation plot (\cref{fig:bifurcation}). We varied the parameter $\phi_u$, while keeping other parameters at the baseline values to calculate the $\bbR_0$ values and trace out different steady states. The stability of the steady states leads to the backward bifurcation behavior, which highlights a critical threshold condition for establishing \Ws infection among mosquitoes over a range of $\bbR_0$ values. For example, for a given $\bbR_0$, we identify the minimum fraction of infection that needs to be exceeded in females to establish a stable endemic state of \textit{Wolbachia}. Here, we are assuming a natural distribution of infection among the population. Above this threshold, the system will approach the complete-infection stable equilibrium. Below the threshold, the system will approach the disease-free stable equilibrium. This threshold infection rate is $34\%$ among females for the baseline case, where $\bbR_0=0.68$.


The backward bifurcation behavior can be interpreted as the result of the competition between the infected and uninfected mosquito cohorts. Recall that $\bbR_0$ is defined as the ratio between the next-generation factors for uninfected $\bbG_{0u}$ and infected $\bbG_{0u}$. We will describe two regimes as a competition between the cohorts of infected and uninfected mosquitoes.

When $\bbR_0>1$ ($\bbG_{0w}>\bbG_{0u}$), there is an unstable DFE, which means that the introduction of one infected mosquito will cause a rapid spread of \textit{Wolbachia} among the population. However, under the biologically-relevant regime, since the \Ws infection affects the fitness of the infected mosquitoes, the reproduction $\bbG_{0w}<\bbG_{0u}$ or $\bbR_0\leq1$, and the introduction of a small fraction of infection at DFE will die out. This scenario is consistent with the fact that \Ws infection is not naturally found in wild \textit{Anopheles} mosquitoes.

Because infected males sterilize the uninfected females, if we increase the fraction of infected mosquitoes, the cytoplasmic incompatibility decreases the fitness of the uninfected mosquitoes. As more infected mosquitoes are released, the fitness of the uninfected mosquitoes becomes less than the infected mosquitoes. The fraction where the two fitnesses are the same corresponds to the threshold condition for sustaining the endemic \Ws infection. 

\section{Numerical Simulations}\label{sec:numerical}
Our numerical simulations aim to provide qualitative insights for designing optimal release strategies to establish a stable \Ws infection in mosquito populations by comparing practical field release programs. We compare pre-release larvicide and thermal fogging, releasing multiple batches of \Ws mosquitoes and the time of year for the release.

\subsection{Parameter Estimations}\label{sec:param}
Most of our estimates are based on the work by Joshi et al. \cite{joshi2014wolbachia}, which characterize the life parameters of the mosquitoes in an ideal lab setting. These parameters can vary depending on environmental factors. In \cref{sec:seasonality}, we discuss the impact of seasonality on \W-releasing strategies.

\paragraph{\textbf{Maternal transmission}}
Maternal transmission from infected females to their offspring is the primary mechanism that \Ws is transmitted to other mosquitoes. \textit{wAlbB}-infected females have close to perfect (100\%) maternal transmission where almost all of the offspring of infected females are infected. 

\paragraph{\textbf{Mosquito lifespan}}
\W-infection has a minor, if any,  impact on the longevity of males. The median lifespan, under the laboratory condition, is about 16 days \cite{joshi2014wolbachia}. However, in the competitive field environment, the lifespan is shorter \cite{CDCMalaria}; thus, we use the realistic estimate for the lifespan of male mosquitoes ($\tau_{mu}=\tau_{mw}=7$ days). When fed on human blood, infected and uninfected females live for approximately 22 days in the lab. However, the infected females show better survivorship during the first two weeks \cite{joshi2014wolbachia}. Thus, we assume a shorter lifespan in the realistic setting, $\tau_{fu} = 13$ days and $\tau_{fw} = 15$ days.

\paragraph{\textbf{Egg-laying rates}}
\Ws infection has a negligible impact on the total number of eggs an \An female mosquito produces throughout a lifetime, which is about $50$ eggs/female \cite{joshi2014wolbachia}. Thus, the daily egg-laying rates for the uninfected females ($\phi_u$) is $50/\tau_{fu} \approx 3.8$ eggs/day, and $50/\tau_{fw} \approx 3.3$ eggs/day for the infected females ($\phi_w$).

\paragraph{\textbf{Egg-hatching rate and death rates}}
We assume that the \Ws infection does not impact the egg-hatching period. On average, it takes about three days for eggs to hatch \cite{CDCMalaria}; thus, we set $\tau_\delta = 3$ days and $\delta=1/\tau_{\delta}=1/3$. \Ws infection reduces the fecundity in female mosquitoes; the fraction of eggs that hatch and survive to first-instar larvae is lower among the infected population than the uninfected one ($50\%$ vs. $73\%$) \cite{joshi2014wolbachia}. Thus, we have $\delta/(\delta+\mu_{eu})=0.73$ and $\delta/(\delta+\mu_{ew})=0.5$, which yield death rate estimates of $\mu_{eu}\approx 0.12$ and $\mu_{ew}\approx 0.33$ for eggs.

\paragraph{\textbf{Larvae/pupae emergence rate and death rate}}
\Ws has no significant impact on the life traits of \An mosquitoes, including the emergence time from larvae to adults and survivorship. On average, it takes about $\tau_\psi =18$ days (sum of pupation time and emergence time, see \cite{joshi2014wolbachia}), which gives the emergence rate $\psi=1/\tau_\psi=1/18$. The fraction of larvae that survive to the adult stage is $\psi/(\psi+\mu_l) \approx 80\%$; thus, the daily death rate for the larvae/pupae stage is estimated as $\mu_l = 0.01$.

\subsection{Sensitivity Analysis}
We quantify the significance of the parameters in the model predictions using a local sensitivity analysis. This helps us better understand our model when parameters are changed. 

We use the normalized relative sensitivity index of a quantity of interest (QOI), $q(p)$, with respect to a parameter of interest $POI$, $p$, defined as 
$\mathcal{S}_{p}^{q}:=\frac{p}{q}\times\frac{\partial q}{\partial p}$. This index measures the percentage change in the QOI given the percentage change in the input POI. In other words, if parameter $p$ changes by $\alpha\%$, then $q$ will change by $\mathcal{S}_{p}^{q}\times\alpha\%$. The sign determines the decreasing or increasing behavior of the quantity. We evaluate the index at the baseline parameter values to obtain the local sensitivity index.

\begin{table}[ht]
\centering
\caption{Sensitivity Analysis. Sensitivity indices for threshold-related quantities (in the first column) with respect to the model parameters (first row). Threshold (row 3) refers to the threshold level of infection and the time (row 4) measures the time to achieve $90\%$ infection.}\label{tab:SA}
\setlength{\tabcolsep}{1pt}
{\small
\begin{tabular}{cSSSSSSSSSSSSS}
\toprule
& $\nu_w$ & $\phi_w$ & $\phi_u$ & $\mu_{fw}$ & $\mu_{fu}$ & $\mu_{ew}$ & $\mu_{eu}$ & $\delta$ & $\psi$ & $\mu_{mu}$ & $\mu_{mw}$ & $\mu_l$ & $\mu_{adults}$ \\
\midrule
$\bbR_0$ & 1 & 1 & -1 & -1 & 1 & -0.5 & 0.26 & 0.23 & 0 & 0 & 0 & 0 & \num{-2.4e-13} \\
Threshold & -3.5 & -2.1 & 2.1 & 1.4 & -1.4 & 1 & -0.55 & -0.48 & 0 & 0 & 0 & 0 & \num{4e-13} \\ 
Time & -6.7 & -1.1 & 1.1 & 0.88 & -0.6 & 0.51 & -0.32 & -0.32 & -0.6 & -0.69 & 0.6 & -0.14 & 0.14  \\
\bottomrule
\end{tabular}
}
\end{table}

We consider three different QOIs concerning the establishment of \Ws infection: the reproductive number $\mathbb{R}_0$; the threshold of infection in females, which corresponds to the unstable equilibrium indicated in the bifurcation diagram of \cref{fig:bifurcation}; and the establishment time, measured as the time to achieve $90\%$ infection for a particular release setting of interest (\cref{fig:releases}, with pre-release mitigation and released in 5 batches). 

The sensitivity indices are ranked by magnitude (importance) for the QOI = threshold case in \cref{tab:SA}. Following this criterion, the maternal transmission rate $\nu_w$ is the most sensitive parameter among all the selected POIs. In addition, the egg-laying rates ($\phi_u$ and $\phi_w$) and the adult mosquito lifespans of females ($\mu_{fw}$ and $\mu_{fu}$) have also a significant sensitivity with respect to the QOIs. That is, the parameters related to the reproduction and CI of mosquitoes are critical to both the threshold and the speed of establishing a sustained \Ws infection.

Conversely, parameters involved in the survival of eggs, like the egg lifespans ($\mu_{eu}$ and $\mu_{ew}$) and their hatching rate $\delta$, are less sensitive in the three QOIs studied. Furthermore, due to the assumption $\mu_{eu}=\mu_{ew}$, the adult male lifespans ($\mu_{mu}$ and $\mu_{mw}$) do not represent an impact in $\mathbb{R}_0$ and the threshold condition. Instead, they play a relevant role in the time to establish a sustained infection once infection exceeds the threshold.

We also simultaneously perturb all the adult death rates  $\mu_{fu}$, $\mu_{fw}$, $\mu_{mu}$, and $\mu_{mw}$. This simulates a change in the global environment that affects both infected and uninfected mosquitoes. As indicated by the $\mu_{adults}$ column, this change does not have a significant impact on $\bbR_0$ and threshold. This can be understood by checking the individual sensitivity indices for $\mu_{fw}$ and $\mu_{fu}$, the perturbation of which give the same amount of change in opposite directions. Thus, the simultaneous change neutralizes the impact and does not affect the competition outcome. Meanwhile, the change does delay the establishment process as the infected cohort ($\mu_{fw}, \mu_{mw}$) has a larger impact on the establishment time.

\subsection{Compare Pre-release Mitigation Strategies}
To reduce the number of infected mosquitoes released and more efficiently establish a stable \Ws infection, integrated control strategies are often implemented to reduce the wild mosquito population before releasing the infected mosquitoes. We will evaluate the establishment of \Ws when combining pre-release mitigation approaches, including larviciding and spraying. Not all of those strategies are primary vector control interventions in Haiti. Nevertheless, our results inform the potential effectiveness should such interventions become prevalent in the area.

\textit{Larviciding} treats mosquito breeding sites with bacterial or chemical insecticides to kill the aquatic stage of mosquitoes. Field studies of bacterial larvicide products, targeting \An larval habitats, report larval reduction between 47\% and 100\% \cite{derua2019bacterial}. Our model simulated a range of mitigation efficacy (in reducing population) from a more challenging setting of 0.2 to a high efficacy of 0.6.

\textit{Space spraying} or \textit{fogging} refers to dispersing a liquid fog of insecticide into an outdoor area to kill adult insects. The insecticide may be delivered using hand-held, vehicle-mounted, or aircraft-mounted equipment \cite{world2003space}.
The impact of fogging as a malaria vector control intervention for reducing adult \An mosquitoes fluctuates between $50\% \sim 100\%$ \cite{Pryce_Choi_Richardson_Malone_2018}, and we evaluated the impact of space spraying for a moderate range of efficacy, where the mitigation efficacy from $0.2 \sim 0.6$.

As summarized in \cref{tab:PRM}, starting from the baseline DFE state, we simulate the pre-release mitigation (column 1) at different intensities by adjusting the DFE according to the mitigation efficacy at the targeting stage(s) (column 2 - 3). We have assumed that the pre-release mitigations only impact the wild mosquitoes and not the released infected mosquitoes. We then released an equal number of infected males and females, and we identified the threshold quantity (needed for establishing the \Ws endemic) without a time limit (column 4) and with a time limit of two months (column 5). The release size is quantified using the \emph{release factor}, which is the ratio between the number of released mosquitoes and the females at DFE, i.e., $F_u^0$ in \cref{eq:DFE}.

\begin{table}[htbp]
\centering
\caption{\textbf{Comparison of pre-release mitigation strategies target different mosquito life stages (larvae and adults)}. ``Mitigation efficacy'' measures the fraction of population reduced given the mitigation approach, and the ``release factor'' measures the release size of the infected population relative to the baseline female population size at DFE ($F_u^0$). Threshold release sizes needed to establish \Ws within (two months) or without time constraints are identified.}
\begin{tabular}{p{3cm}p{1.5cm}p{1.5cm}p{1.5cm}p{3cm}}
\toprule
Pre-release Mitigation  & Larvae \newline mitigation efficacy & Adults \newline mitigation efficacy & Threshold \newline release factor & Release factor to reach $90\%$ by two months \\ 
\midrule
No mitigation (DFE) &  0 & 0 &  1.13 &  9.9\\
\midrule
\multirow{3}{2cm}{Spraying} &  0 & 0.2 & 1.03 & 9.2 \\ 
&  0 & 0.4 &  0.93 & 9.2\\ 
&  0 & 0.6  &  0.82 &  7.9\\ 
\midrule
\multirow{3}{2cm}{Larviciding} & 0.2 & 0 & 1.04 &   8.0 \\ 
 & 0.4  & 0 & 0.96 & 6.5 \\ 
 & 0.6  & 0 & 0.88 & 5.3 \\  
\midrule
\multirow{3}{2cm}{Spraying + Larviciding}  & 0.6 & 0.2 & 0.79 & 4.8 \\ 
 & 0.6 & 0.4 & 0.69 & 4.4 \\ 
 & 0.6 & 0.6 & 0.60 & 3.9\\  
\bottomrule
\end{tabular}\label{tab:PRM}
\end{table}  

\begin{figure}[htbp]
\centering
\includegraphics[width=\textwidth]{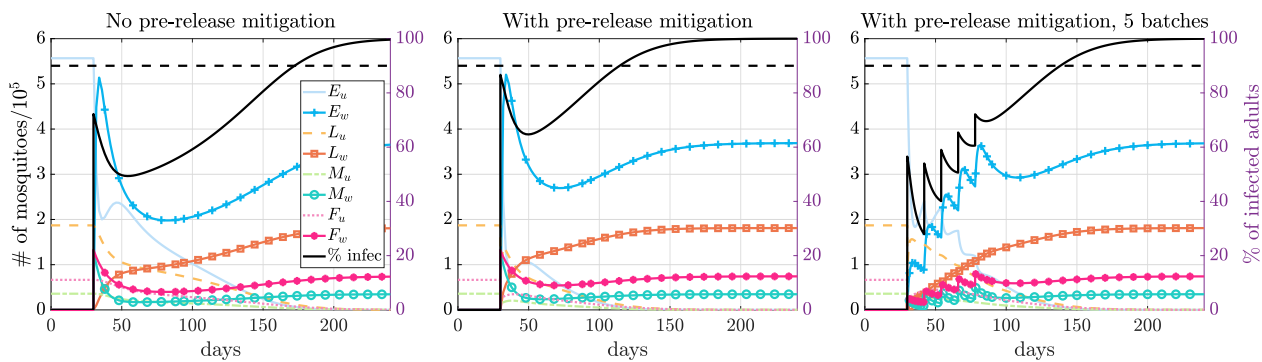}
\caption{\textbf{Simulations for different release scenarios.} An equal number of infected male and female mosquitoes are released (release size = 2) without or with pre-release mitigation (reduced to 40\% in both larvae and adults using hybrid fogging and larviciding, see \cref{tab:PRM}). The black line is the percent of infected mosquitoes that are infected as they are released in one batch or multiple patches. The \Ws endemic is established, and the infection reaches $90\%$ infection around 143, 85, and 109 days after the initial release.}
\label{fig:releases}
\end{figure}

With pre-release mitigation, larger mitigation efficacy reduces the release factor for both spraying and larviciding. This is expected since the threshold is determined by the competition (or ratio) between the infected and uninfected cohorts. Fewer infected mosquitoes are needed to match the competition if there are fewer uninfected mosquitoes in the field.

Under the same release size, pre-release mitigation helps to speed up the establishment of \Ws endemic (\cref{fig:releases}). We also see that spraying requires a slightly smaller release size than larviciding under the same intervention intensity. When applying two interventions together, it outperforms the individual case as expected. 
When releasing just above the threshold, it may take a long time to establish \Ws endemic. The identified threshold value may not be practical due to various model assumptions, such as seasonality. When imposing a two-month time limit, many more infected mosquitoes must be released. Such a large release size may not be practical for field trial implementation; thus, we study the multiple-release strategy next.

\subsection{Multiple Releases}
\begin{figure}[htbp]
\centering
\includegraphics[width=0.5\textwidth]{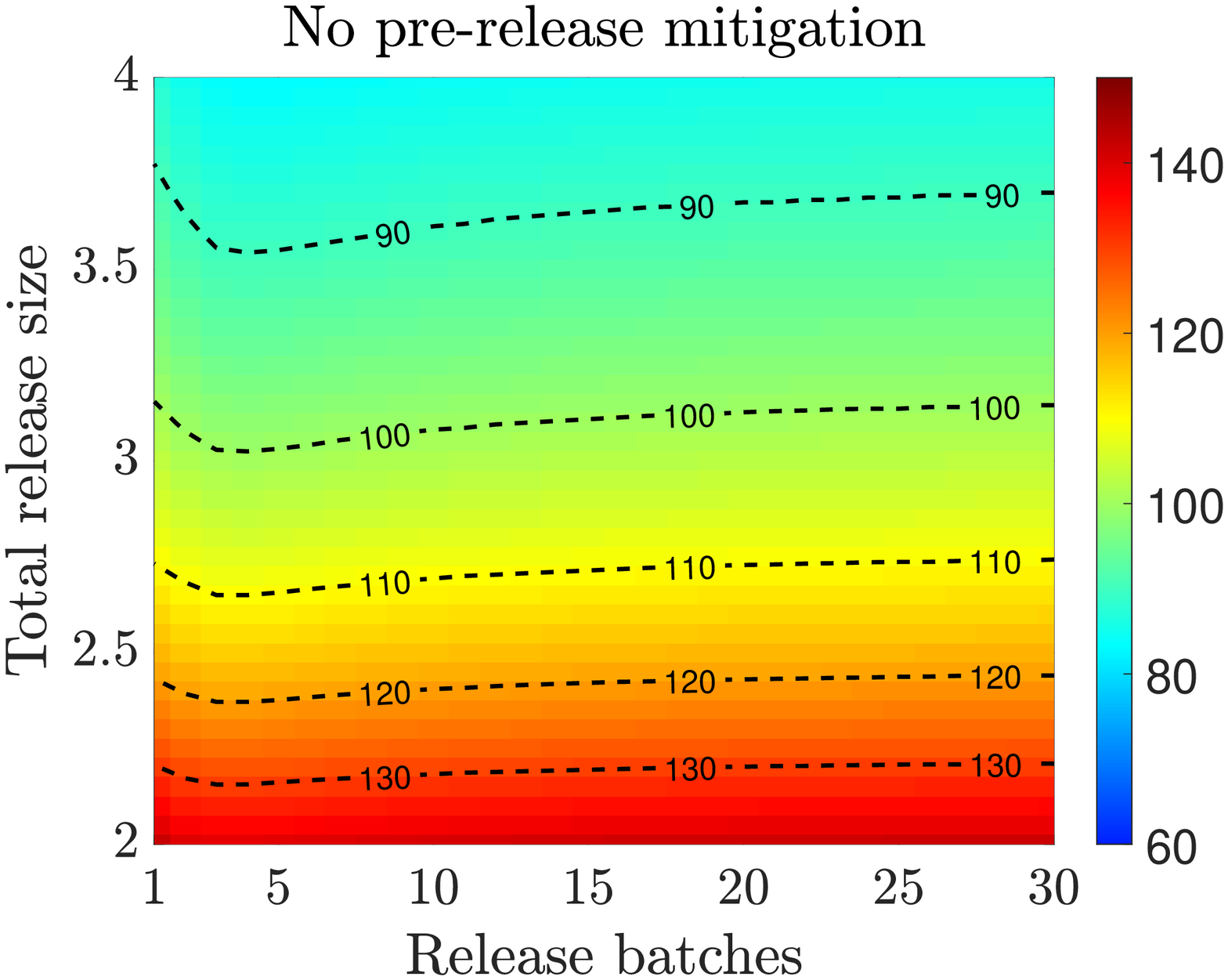}\hfill\includegraphics[width=0.5\textwidth]{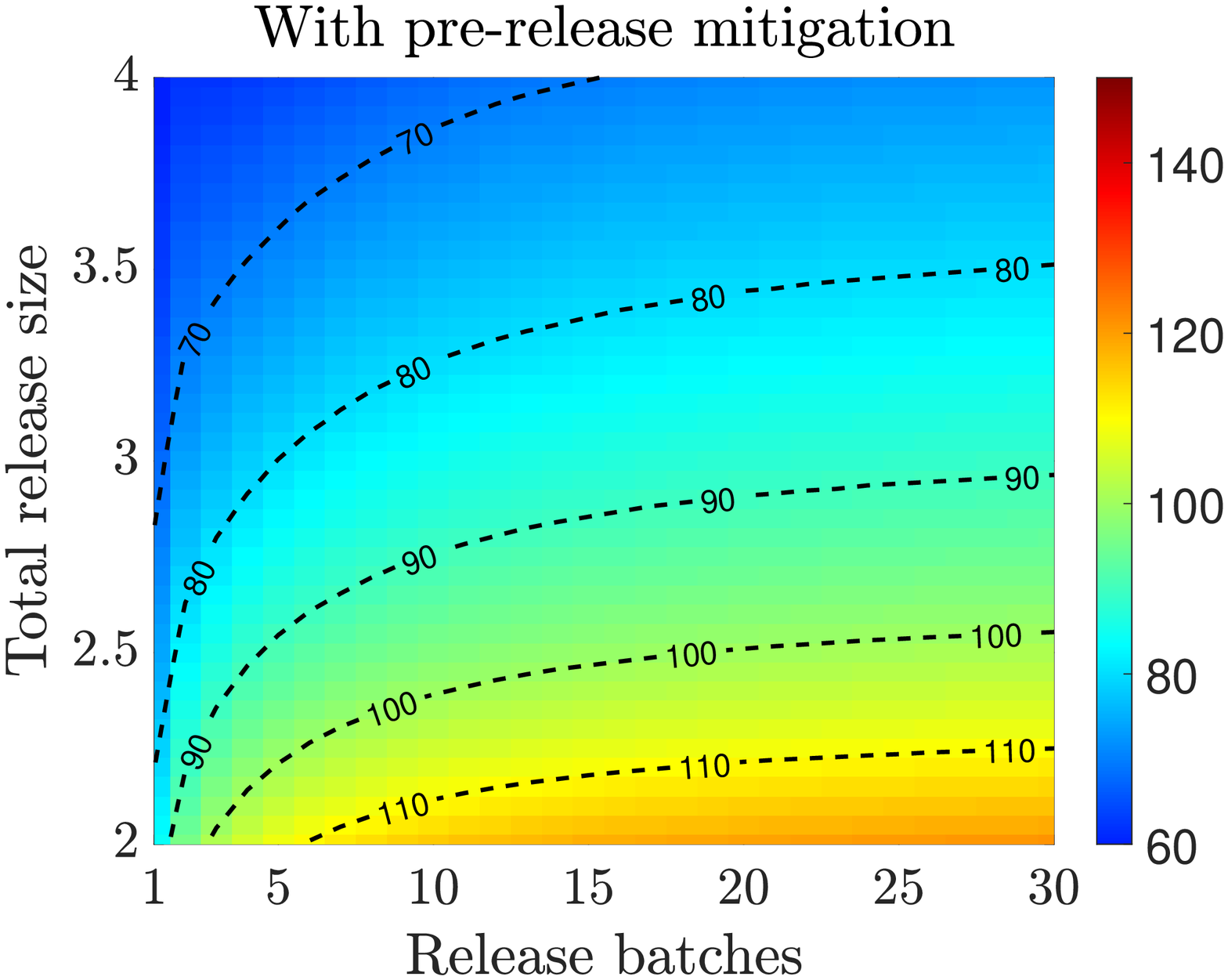}
\caption{\textbf{Compare multiple-release strategy for \Ws establishment speed} with and without pre-release mitigation. The heatmaps indicate the days to achieve $90\%$ infection (color-coded according to the respective color bars). An optimal number of release batches is observed for large releases without pre-release mitigation. Releasing all infections at once is more efficient than splitting them into multiple batches when implementing pre-release mitigation.}
\label{fig:multiple_release}
\end{figure}

Field trials often require periodic releases of batches of infected mosquitoes. We aim to inform an optimal design of multiple release strategy. We consider a certain amount of mosquitoes (release size), split into multiple batches (release batches) and released over two months. We assume all the batches have an equal batch size and equal time gap in between. In \cref{fig:multiple_release}, we plot the time to achieve 90\% infection when using different numbers of release batches and total release sizes, and we study how the establishment time is impacted when using the pre-release mitigation. 

For both scenarios, using a larger release size always helps to speed up the establishment of \Ws infection for both release scenarios. Without pre-release mitigation (\cref{fig:multiple_release} left), the optimal multiple-release strategy is to leave an about two-week gap (four or five batches within two months) between two consecutive releases. The benefit of such a release gap is more significant as the overall release size increases. The necessity of the release gap results from the limited environmental resources available (carrying capacity). 
Releasing all the infections at once may not be as optimal as splitting the release into multiple batches due to the higher penalization from the carrying capacity. Nonetheless, using too many batches decreases invasion efficiency.

On the other hand, when there is pre-release mitigation (\cref{fig:multiple_release} right), it creates a gap in the carrying capacity, which provides an opportunity for instant population replacement by the infected cohort. Thus, releasing infected mosquitoes all at once is more efficient than splitting the release of infection in batches. 

\subsection{Seasonality\label{sec:seasonality}}
Environmental and climactic covariates, such as rainfall and temperature, affect all the stages of the mosquito life cycle. They impact the density and distribution of vector breeding sites, the number of eggs laid, the ability of larvae to emerge from eggs once they are laid (hatching or emergence rate), and the adult mosquito lifespan. Regional carrying capacity is also affected as this parameter is directly influenced by the number of available vector breeding and egg-laying sites. It is important to account for those seasonality effects by adjusting parameters as these values influence the ability to achieve endemic, stable \textit{wAlbB} transmission among the mosquito population. 

We have extracted the CHIRPS monthly data for the department of Grand Anse \citep{Funk2015CHIRPS} in Haiti, where most of the country's malaria transmission occurs. In particular, we considered the seasonality pattern based on the rainfall, humidity, and temperature data. We include a summary of the data we used in  \cref{tab:seasonality_data}. 

The monthly rainfall data suggested a bimodal seasonal pattern with the peak rainfall in May and September (\cref{fig:seasonality_fit}). Therefore, we adapt to a time-dependent carrying capacity, $K_l(t)$, which varies according to a fitted seasonality curve based on the rainfall data. We also simulate release scenarios starting in the dry or rainy season. 
There is a similar seasonal trend in the humidity data (\cref{fig:seasonality_fit}), measured by the aridity index, with most of the year classified as humid. We captured the impact of humidity by using the same time-varying carrying capacity curve above, and we assumed that it does not impact other life traits of mosquitoes.

The monthly temperature data ranged from 25.7 to 29.8 degrees Celsius (78.2  to 85.7 degrees Fahrenheit). Whereas temperature can influence egg laying rates, larval emergence rates, and adult mosquito lifespan, mean monthly temperature in our region of focus does not vary enough to influence rates for these parameters in our model \cite{christiansen2014temperature,beck2013effect}.

\begin{figure}[ht]
\centering
\includegraphics[width=0.48\textwidth]{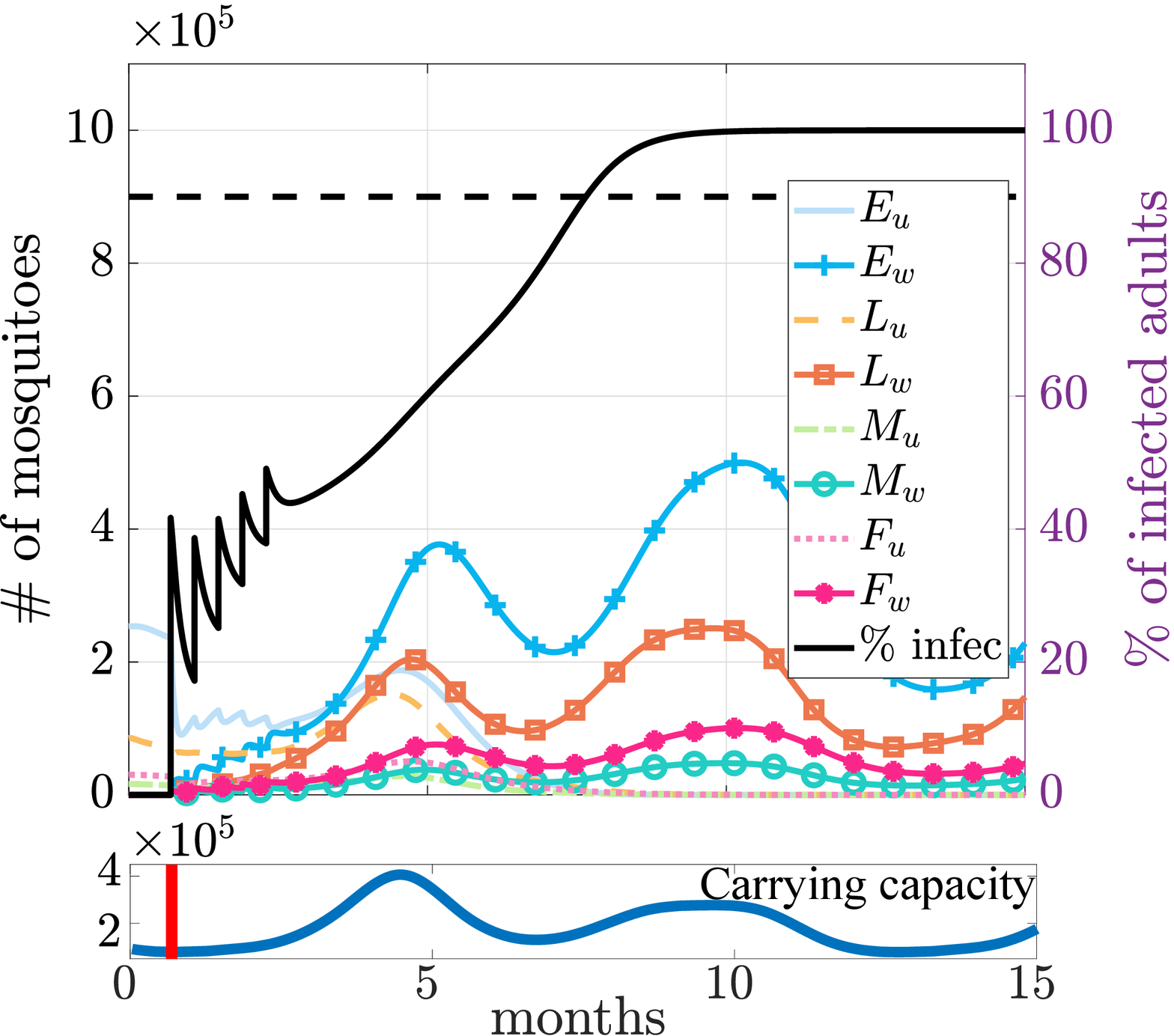}\hfill\includegraphics[width=0.48\textwidth]{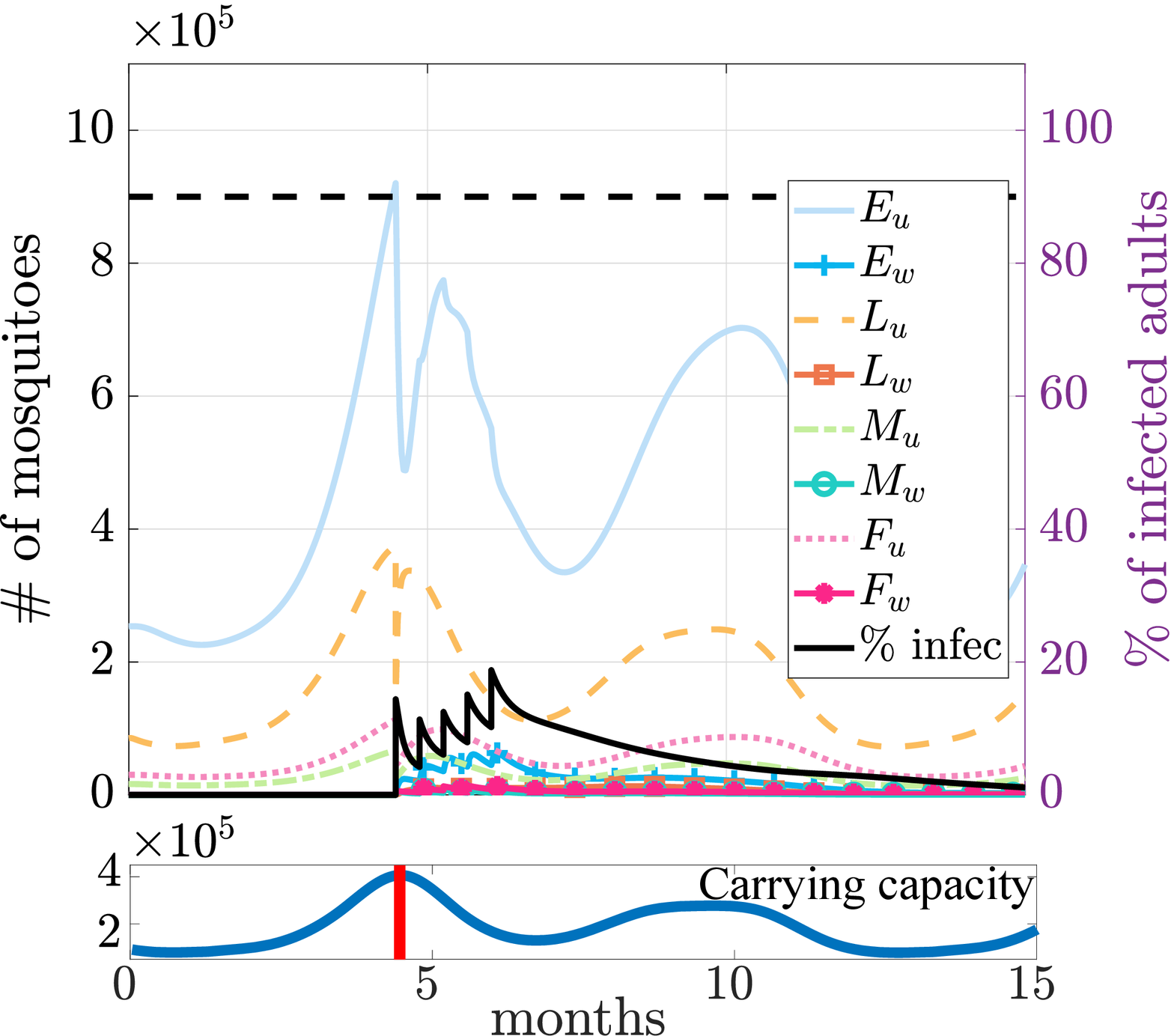}
\caption{\textbf{Impact of seasonality on the field release.} Simulations of releasing the same number of infected mosquitoes at the driest (day = 21, left plot) vs. the wettest (day = 134, right plot) time of the year, indicated by the red vertical bars on the corresponding lower panels. The black line for the percent of infected mosquitoes shows that the infection is successfully established when released in the drier time, while it dies out when released in the wetter time.}
\label{fig:season}
\end{figure}

We aim to study seasonality's impact on field releases' efficacy in establishing a \Ws infection. For this purpose, we considered releasing an equal number of female and male infected mosquitoes with release factor 1 (as defined in \cref{tab:PRM}), relative to the uninfected female population at the DFE on day 1 of the year, i.e., $F_u^{0}=30270$). In addition, the total quantity was released in five batches. 

Under the above setting, our simulation results suggest that it is more efficient to establish \Ws during the dry season. Releasing infection during the dry season (\cref{fig:season} left) requires releasing fewer infected mosquitoes to exceed the threshold. In contrast, when releasing the same amount of mosquitoes during the wet season (\cref{fig:season} right), the infection fails to establish due to the abundance of wild mosquitoes (higher carrying capacity).

\section{Discussion}
We developed and analyzed a compartmental ODE model to describe the establishment of \Ws infection in wild \An mosquitoes. The model tracks male and female mosquitoes through the egg, larval, and adult stages. The model accounts for maternal transmission of \W, cytoplasmic incompatibility, and fitness cost induced from \Ws infection. Also, we incorporated carrying capacity constraints on mosquito population size to study the impact of seasonality, specific to Haiti. 

The next-generation numbers, $\mathbb{G}_{0u}$ and $\mathbb{G}_{0w}$, were derived to estimate the number of new offspring reproduced per generation for the uninfected and infected mosquitoes. The ratio of the two factors gives the basic reproductive number $\bbR_0$, which characterizes the initial invasion dynamics of the \W infection. At baseline, $\bbR_0<1$ suggests that the infection will die out by releasing a small number of \Ws- infected mosquitoes.

Using stability analysis, we identified a threshold condition for establishing a stable \Ws infection. The system demonstrates a backward bifurcation separating an unstable disease-free equilibrium and a stable complete-infection equilibrium with an unstable endemic equilibrium. 

We simulated \Ws releases with pre-release mitigations, including thermal fogging and larvicide. We measured the efficacy of these different strategies by the number of \W-infected mosquitoes needed to exceed the threshold condition and time (in days) it took to reach an endemic equilibrium of \Ws. Our simulations show that the pre-release mitigations can lower the threshold and accelerate the establishment of \Ws infection. In the absence of pre-release mitigations, the establishment of \Ws infection is achieved at a later time.

We studied seasonality's impact by varying mosquito populations' carrying capacity. Our results indicate that releasing \W-infected mosquitoes in the dry season is more effective than in the rainy season when fewer uninfected mosquitoes are present in the wild for competition. 

We studied the potential benefit of conducting multiple releases of infected mosquitoes, given the limitation of production capacity in raising \W-infected mosquitoes in field settings. With the pre-release mitigations, we found that multiple releases may take more time to reach an endemic equilibrium of \Ws when compared to a single release.

The relative importance of the model parameters
were ranked using sensitivity analysis. We found that the maternal transmission rate is the most sensitive parameter. This varies widely between different species of \Ws (\cref{tab:ACI}) and must be considered when applying our results to different species. The results are also very sensitive to the egg-laying rates and lifespans of adult females.

Mathematical models simplify field conditions; thus, the results from models should be judiciously interpreted. For example, we accounted for the seasonality by varying the carrying capacity. While this is an appropriate approximation for the mild variation of temperature and humidity in Haiti, a more complex model would be necessary for studying the locations where seasonality is more prominent. 

Another major limitation is our model parametrization, which results from the sparse publication of parameter values for \W-infection in African \An vectors. For example, we assumed the death rate of uninfected females is slightly higher than \W-infected females based on \cite{joshi2014wolbachia}. Additional studies from lab and field settings are needed to confirm the fitness cost of \Ws infection in African \An species. Also, our modeling results are only valid for the \textit{wAlbB} strain of \W. Here we assumed perfect maternal transmission and \W-induced CI. Thus, parameter values would differ for other strains, where these two assumptions do not hold.

We analyzed baseline mitigation strategies to optimize sustaining an endemic \Ws infection, specifically among wild mosquito populations. These approaches do not reflect all the current interventions in Grand'Anse, Haiti. 
Other vector-borne malaria interventions were not considered, such as indoor residual spraying and insecticide-treated nets. 
These need to be considered in future studies when estimating the impact of the \W-based population replacement strategy on malaria transmission among human populations.

Our ODE model assumes that the fraction of infection among mosquitoes is homogeneous in space. However, this may not hold in field settings, especially when modeling a local release of infected mosquitoes. Therefore, it is essential to include the impact of spatial dynamics to determine the threshold condition for field releases. We are developing a partial differential equation (PDE) model to study the invasion dynamics of \Ws infection among mosquitoes in a more realistic field setting. This reaction-diffusion-type model accounts for complex maternal transmission and spatial mosquito dispersion. Our initial studies have identified an optimal bubble-shaped distribution to minimize the number of mosquitoes needed to exceed the threshold conditions \cite{qu2022modeling}.

\appendix
\section[\appendixname~\thesection]{Proof of Theorem 1 (Stability of DFE) }\label{sec:A1}
Let us consider the variable vector containing the compartments (re-arranged by the infection status), $\textbf{Y}=(E_u, L_u, F_u, M_u, E_w, L_w, F_w, M_w)$, then the corresponding Jacobian, denoted by $J$, of the system $\frac{d\textbf{Y}}{dt}=J\textbf{Y}$, is given by\\

\scalemath{0.85}{\hspace{-0.8cm}\mathbf{J}=
\left(
\begin{array}{cccc;{2pt/2pt}cccc}
-\delta-\mu_{eu} & 0 & \mathbb{M}_u\phi_u & 0 & 0 & 0 & v_u \phi_w & 0 \\
\delta\mathbb{K} & -\frac{\delta E_u+K_l(\mu_l+\psi)}{K_l} & 0 & 0 & 0 & -\frac{\delta}{K_l}E_u & 0 & 0 \\
0 & b_f\psi & -\mu_{fu} & 0 & 0 & 0 & 0 & 0\\
0 & b_m\psi & 0 & -\mu_{mu} & 0 & 0 & 0 & 0\\ \hdashline[2pt/2pt]
0 & 0 & 0 & 0 & -\delta-\mu_{ew} & 0 & v_w\phi_w & 0\\
0 & -\frac{\delta}{K_l}E_w & 0 & 0 & \delta\mathbb{K} & -\frac{\delta E_w+K_l(\mu_l+\psi)}{K_l} & 0 & 0\\
0 & 0 & 0 & 0 & 0 & b_f\psi & -\mu_{fw} & 0\\
0 & 0 & 0 & 0 & 0 & b_m\psi & 0 & -\mu_{mw}
\end{array} 
\right)~.
}~~\\

At the disease-free equilibrium, the Jacobian becomes
\begin{equation}\label{eq:JDFE}
\mathbf{J}_{DFE}=\left(
\begin{array}{c;{2pt/2pt}c}
\mathbf{A}_{DFE} & * \\ \hdashline[2pt/2pt]
\mathbf{0} & \mathbf{D}_{DFE}
\end{array} 
\right)~, \quad \text{where}   
\end{equation}

\begin{equation}\label{eq:ADFE}
\mathbf{A}_{DFE}\!=\!\!\left(
\begin{array}{ccc;{2pt/2pt}c}
-\delta-\mu_{eu} & 0 & \phi_u & 0\\
\frac{\delta}{\mathbb{G}_{0u}} &  -\mathbb{G}_{0u}(\mu_l+\psi) & 0 & 0\\
0 & b_f\psi & -\mu_{fu} & 0\\\hdashline[2pt/2pt]
0 & b_m\psi & 0 & -\mu_{mu}
\end{array} 
\right)=
\left(
\begin{array}{c;{2pt/2pt}c}
A_{s1} & \mathbf{0} \\ \hdashline[2pt/2pt]
* & -\mu_{mu}
\end{array} 
\right)~,
\end{equation}
and 
\begin{equation}\label{eq:DDFE}
\mathbf{D}_{DFE}=\left(
\begin{array}{ccc;{2pt/2pt}c}
-\delta-\mu_{ew} & 0 & v_w\phi_w & 0 \\
\frac{\delta}{\mathbb{G}_{0u}} & -\mu_{l}-\psi & 0 & 0 \\ 
0 & b_f\psi & -\mu_{fw} & 0\\ \hdashline[2pt/2pt]
0 & b_m\psi & 0 & -\mu_{mw}
\end{array} 
\right)=\left(
\begin{array}{c;{2pt/2pt}c}
D_{s1} & \mathbf{0} \\ \hdashline[2pt/2pt]
* & -\mu_{mw}
\end{array} 
\right)~.
\end{equation}
Observe first that the matrix $\mathbf{J}_{DFE}$ is an upper triangular block matrix, and the two $4\times4$ diagonal elements $\mathbf{A}_{DFE}$ and $\mathbf{D}_{DFE}$ (\cref{eq:JDFE}). Thus, the stability of the matrix $\mathbf{J}_{DFE}$ is equivalent to showing the stability for both matrices.

To show the stability of $\mathbf{A}_{DFE}$, notice that it's a lower triangular block matrix with $-\mu_{mu}<0$ (\cref{eq:ADFE}), thus we just need to consider the $3\times3$ leading principal submatrix $A_{s1}$, which can be further partitioned as follows,
\begin{equation*}
A_{s1} =
\left(
\begin{array}{cc;{2pt/2pt}c}
-\delta-\mu_{eu} & 0 & \phi_u\\
\delta/\mathbb{G}_{0u} & -\mathbb{G}_{0u}(\mu_l+\psi) & 0\\ \hdashline[2pt/2pt]
0 & b_f\psi & -\mu_{fu} 
\end{array}
\right)
=\left(
\begin{array}{c;{2pt/2pt}c}
A_1 & B_1\\\hdashline[2pt/2pt]
C_1 & D_1
\end{array}
\right)~.
\end{equation*}
To prove the stability of the matrix $A_{s1}$, we use a result on Metzler matrices stated in Proposition 3.1 in \cite{kamgang2008computation}. Since $A_{s1}$ is a Metzler matrix, $A_{s1}$ is Metzler stable if and only if $A_1$ and $D_1-C_1A_1^{-1}B_1$ are Metzler stable. With this in mind, observe that since the matrix $A_1$ is lower triangular, its eigenvalues are its corresponding diagonal entries $-(\delta+\mu_{ew} )$ and $-\mathbb{G}_{0u}(\mu_l+\psi)$, which are both negative. This implies that $A_1$ is Metzler stable. Meanwhile, the matrix $D_1-C_1A_1^{-1}B_1$ satisfies the property:
\begin{equation}\label{eq:cond1}
D_1-C_1A_1^{-1}B_1= \mu_{fu}\Big(\frac{1}{\mathbb{G}_{0u}}-1\Big)< 0,\;\;\text{if}\;\; \mathbb{G}_{0u}>1~.
\end{equation}
Thus, when $\mathbb{G}_{0u}>1$, $A_{s1}$ is Meltzer stable, so as the matrix $\mathbf{A}_{DFE}$.

Similarly, we derive the condition for the stability of $\mathbf{D}_{DFE}$. Given it is a lower triangular block matrix with $-\mu_{mw}<0$ (\cref{eq:DDFE}), we are left with the $3\times3$ leading submatrix
\begin{equation*}
D_{s1} =
\left(
\begin{array}{cc;{2pt/2pt}c}
-\delta-\mu_{ew} & 0 & v_w\phi_w\\
\delta/\mathbb{G}_{0u} & -(\mu_l+\psi) & 0\\ \hdashline[2pt/2pt]
0 & b_f\psi & -\mu_{fw} 
\end{array}
\right)
=\left(
\begin{array}{c;{2pt/2pt}c}
A_2 & B_2\\\hdashline[2pt/2pt]
C_2 & D_2
\end{array}
\right)~. 
\end{equation*}
By the same argument used previously, $A_2$ is Metzler stable and $D_2-C_2A_2^{-1}B_2$ satisfies the inequality:
\begin{equation}\label{eq:cond2}
D_2-C_2A_2^{-1}B_2= \mu_{fw}(\mathbb{R}_0-1)< 0,\;\;\text{if}\;\; \mathbb{R}_{0}<1~.
\end{equation}
In consequence, when $\mathbb{R}_{0}<1$, $D_{s1}$ is Metzler stable, so as the matrix $\mathbf{D}_{DFE}$.

Finally, combining the two conditions \cref{eq:cond1,eq:cond2} for the stability of $\mathbf{A}_{DFE}$ and $\mathbf{D}_{DFE}$, we conclude that $\mathbf{J}_{DFE}$ is stable provided that $\mathbb{G}_{0u}>1$ and $\mathbb{R}_{0}<1$. \qed

\section[\appendixname~\thesection]{Proof of Theorem 2 (Stability of CIE) }\label{sec:A2}
At the complete infection equilibrium, the Jacobian has the form:

\begin{equation*}
\scalemath{0.75}{\hspace{-0.8cm}\mathbf{J_{CIE}}=
\left(
\begin{array}{cccc;{2pt/2pt}cccc}
-\delta-\mu_{eu} & 0 & 0 & 0 & 0 & 0 & 0 & 0 \\
\frac{\delta}{\mathbb{G}_{0w}} & -(\mu_l+\psi)& 0 & 0 & 0 & 0 & 0 & 0 \\
0 & b_f\psi & -\mu_{fu} & 0 & 0 & 0 & 0 & 0\\ 
0 & b_m\psi & 0 & -\mu_{mu} & 0 & 0 & 0 & 0\\ \hdashline[2pt/2pt]
0 & 0 & 0 & 0 & -\delta-\mu_{ew} & 0 & \phi_w & 0\\
0 & -\frac{G_{0w}}{\delta}(\psi+\mu_{l}) & 0 & 0 & \frac{\delta}{\mathbb{G}_{0w}}&  -\mathbb{G}_{0w}(\mu_l+\psi) & 0 & 0\\
0 & 0 & 0 & 0 & 0 & b_f\psi & -\mu_{fw} & 0\\
0 & 0 & 0 & 0 & 0 & b_m\psi & 0 & -\mu_{mw}
\end{array}
\right)~.
}
\end{equation*}
By applying the same argument we used in the previous proof, we consider the $4\times4$ diagonal elements:

\begin{equation*}
\mathbf{A_{CIE}}=\begin{pmatrix}
-\delta-\mu_{eu} & 0 & 0 & 0\\
\frac{\delta}{\mathbb{G}_{0u}} &  -(\mu_l+\psi) & 0 & 0\\
0 & b_f\psi & -\mu_{fu} & 0\\
0 & b_m\psi & 0 & -\mu_{mu}
\end{pmatrix}
\end{equation*}
and 
\begin{equation*}
\mathbf{D_{CIE}}=
\begin{pmatrix}
-\delta-\mu_{ew} & 0 & \phi_w & 0 \\
\frac{\delta}{\mathbb{G}_{0w}} & -\mathbb{G}_{0w}(\mu_{l}+\psi) & 0 & 0 \\
0 & b_f\psi & -\mu_{fw} & 0\\
0 & b_m\psi & 0 & -\mu_{mw}
\end{pmatrix}~.
\end{equation*}
Observe that $\mathbf{A}_{CIE}$ is a Metzler matrix (negative diagonal elements) and it is lower triangular, which means that its eigenvalues are the same negative diagonal elements. Thus, $\mathbf{A}_{CIE}$ is Metzler stable. Now, we need to analyze the stability conditions of $\mathbf{D}_{CIE}$. For this purpose, consider the $3\times3$ leading submatrix of $\mathbf{D}_{CIE}$:

\begin{equation*}
D_{s2} =
\left(
\begin{array}{cc;{2pt/2pt}c}
-\delta-\mu_{ew} & 0 & \phi_w\\
\delta/\mathbb{G}_{0w} & -\mathbb{G}_{0w}(\mu_l+\psi) & 0\\ \hdashline[2pt/2pt]
0 & b_f\psi & -\mu_{fw} 
\end{array}
\right)
=\left(
\begin{array}{c;{2pt/2pt}c}
A_3 & B_3\\\hdashline[2pt/2pt]
C_3 & D_3
\end{array}
\right)~.
\end{equation*}
By applying a similar argument as before, $A_3$ is Metzler stable and the number $D_3-C_2A_2^{-1}B_2$ satisfies the inequality:

\begin{equation}\label{eq:cond3}
D_3-C_3A_3^{-1}B_3= \mu_{fw}\Big(\frac{1}{\mathbb{G}_{0w}}-1\Big)< ~0,\;\;\text{if}\;\; \mathbb{G}_{0w}>1~.
\end{equation}
Hence, when $\mathbb{G}_{0w}>1$, $D_{s2}$ is Metzler stable, so as the matrix $\mathbf{D}_{CIE}$. Therefore, we conclude that \cref{eq:cond3} ensures the stability of $\mathbf{J}_{CIE}$.

\section[\appendixname~\thesection]{Seasonality Fitting}\label{sec:A3}
We extracted CHIRPS monthly rainfall data during 2016-2020 for the department of Grand Anse \citep{Funk2015CHIRPS} in Haiti using R Studio version 4.2.1. The median values for each month over these five years were calculated in \cref{tab:seasonality_data}. Most rainfall occurs during two rainy seasons in Grand Anse, April-June and September-November, and the dry season occurs from December-March and July-August.

\begin{figure}[ht]
\centering
\includegraphics[width=\textwidth]{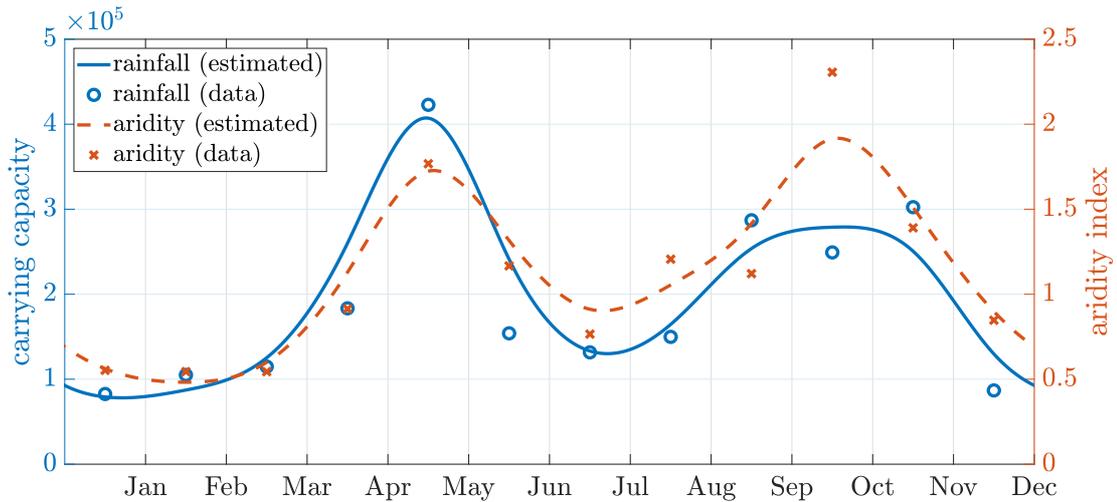}
\caption{Fitted seasonality curves. Left axis: Rescaled rainfall data and the fitted carrying capacity curve (solid blue line). Right axis: Aridity index data and the fitted curve(red dashed line). The raw data values listed in \cref{tab:seasonality_data} and points have been shifted to the center of the month. Both estimated curves are fitted using a periodic smoothing spline, showing a consistent trend.}
\label{fig:seasonality_fit}
\end{figure}

We employed a smoothing spline with a periodicity of 12 months and a smoothing parameter of $0.9$ (implemented in MATLAB) to fit the data points. Assuming that the carrying capacity level is proportional to the amount of rainfall, we rescaled the data points by a constant coefficient ($\approx 1400$) so that the obtained spline predicts an annual average carrying capacity of around $K_l = 2\times10^5$. This is the baseline level assumed in the main text (\cref{tab:param}). The fitted curve and the rescaled data points are shown in \cref{fig:seasonality_fit}.

We also extracted the aridity index data in the region, where only the averaged data for the years 1970-2000 are readily available (\cref{tab:seasonality_data}). The global aridity index is a function of the evapotranspiration processes and rainfall deficit for potential vegetative growth. It is calculated as the mean annual precipitation mean divided by annual reference evapotranspiration.    We employed the similar smoothing spline method described above, and the fitted curve (dashed line in \cref{fig:seasonality_fit}) has a similar trend as the rainfall data. The mean aridity index over the year is $1.13$  with $79\%$ of the time classified as humid (index above $0.65$). Thus, to approximate the carrying capacity, we follow the curve rescaled from the rainfall data.

The temperature variance in the region is mild (see \cref{tab:seasonality_data} row 3). Thus, we have assumed temperature indecency in our model parametrization. 

\begin{table}[ht]
\centering
\begin{tabular}{@{}lllllllllllll@{}}
\hline
&Jan & Feb & Mar & Apr & May & Jun & Jul & Aug & Sep & Oct & Nov & Dec \\ \hline
Rainfall & 59 & 75 & 82 & 131 & 302 & 110 & 94 & 107 & 205 & 178 & 216 & 62 \\
Aridity & 0.6 & 0.5 & 0.5 & 0.9 & 1.8 & 1.2 & 0.8 & 1.2 & 1.1 & 2.3 & 1.4 & 0.8 \\
Temp. & 25.7 & 27.6 & 28.9 & 29.8 &	29.8 & 29.6 & 29.2 & 29.7& 29.3 &	27.8 & 26.4	& 25.7\\
\hline
\end{tabular}
\caption{Seasonality data for the department of Grand Anse in Haiti. Row 1: Monthly median rainfall (in mm) from 2016-2020. Row 2: Mean aridity index for years 1970-2000. Row 3: Mean monthly land surface temperature for 2018-2020 (in Celsius). }
\label{tab:seasonality_data}
\end{table}

\newpage
\bibliographystyle{vancouver2}
\bibliography{bibliography}
  
\end{document}